\documentclass[conference]{acmsiggraph}

\usepackage{pdfpages}

\pdfoutput=1

\TOGonlineid{45678}
\TOGvolume{0}
\TOGnumber{0}
\TOGarticleDOI{1111111.2222222}
\TOGprojectURL{}
\TOGvideoURL{}
\TOGdataURL{}
\TOGcodeURL{}

\usepackage{amsthm}

\makeatletter
\newif\if@restonecol
\makeatother

\usepackage[lined,boxed,commentsnumbered,ruled]{algorithm2e}

\usepackage{epstopdf}

\title{Meshfree $C^2$-Weighting for Shape Deformation}

\author{Chuhua Xian$^{1,2}$~~~~~~Shuo Jin$^{1}$~~~~~~Charlie C. L. Wang$^1$
\\
$^{1}$The Chinese University of Hong Kong~~~~~~$^{2}$South China University of Technology}

\pdfauthor{C. Xian et al.}

\begin{document}

%% \teaser{
%%   \includegraphics[height=1.5in]{images/sampleteaser}
%%   \caption{Spring Training 2009, Peoria, AZ.}
%% }
\teaser{
\includegraphics[width=\linewidth]{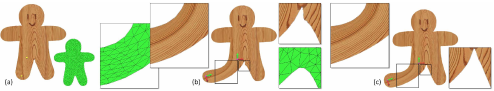}
\centering \caption{Deformation using meshfree $C^2$-weighting.
(a) Handle-driven deformation based on linear blending is an
intuitive method for the interactive shape manipulation. (b)
Artifacts caused by highly distorted triangles can be generated
from the weights computed on an originally well-meshed domain. (c)
We propose a meshfree framework to generate $C^2$-continuous
weights for linear blending based deformation. Our approach
inherits the merits of mesh-dependent weighting schemes meanwhile
bringing the weighting method to the resolution of
infinity.}\label{fig:teaser} }

\maketitle

\begin{abstract}
Handle-driven deformation based on linear blending is widely used
in many applications because of its merits in intuitiveness,
efficiency and easiness of implementation. We provide a meshfree
method to compute the smooth weights of linear blending for shape
deformation. The $C^2$-continuity of weighting is guaranteed by
the carefully formulated basis functions, with which the
computation of weights is in a closed-form. Criteria to ensure the
quality of deformation are preserved by the basis functions after
decomposing the shape domain according to the Voronoi diagram of
handles. The cost of inserting a new handle is only the time to
evaluate the distances from the new handle to all sample points in
the space of deformation. Moreover, a virtual handle insertion
algorithm has been developed to allow users freely placing handles
while preserving the criteria on weights. Experimental examples
for real-time 2D/3D deformations are shown to demonstrate the
effectiveness of this method.
\end{abstract}

\begin{CRcatlist}
\CRcat{I.3.5}{Computer Graphics}{Computational Geometry and Object
Modeling}{Geometric algorithms, languages, and systems}{}
\end{CRcatlist}

\keywordlist

%% Use this only if you're preparing a technical paper to be published in the
%% ACM 'Transactions on Graphics' journal.

\TOGlinkslist

%% Required for all content.

\copyrightspace

%%%%%%%%%%%%%%%%%%%%%%%%%%%%%%%%%%%%%%%%%%%%%%%%%%%%%%%%%%%%%%%%%%%%%%%%%%
%%%%%%%%%%%%%%%%%%%%%%%%%%%%%%%%%%%%%%%%%%%%%%%%%%%%%%%%%%%%%%%%%%%%%%%%%%
\section{Introduction}\label{secIntro}
Shape deformation techniques have various applications in computer
graphics for image manipulation, geometric modeling and animation.
Compared with other deformation strategies, handle-driven methods
outperform others as they are intuitive, effective and
easy-to-implement in many different scenarios. Using handles,
users can bind a shape $\Omega$ with the handles and then
manipulate their locations and orientations to drive the
deformation of $\Omega$. Specifically, each handle $H_i$ with
$i=1,\ldots, m$ is defined as a local frame with its origin
$\mathbf{h}_i \in \Omega$. After defining an affine transformation
$\mathbf{T}_i$ for each handle $H_i$, the deformation of $\Omega$
is realized by computing the new position of each point
$\mathbf{p} \in \Omega$ via a linear blending of affine
transformations $\mathbf{T}_i \mathbf{p}$. The linear blending is
weighted by fields $w_i: \Omega \mapsto \Re$ associated with
handles $H_i$. Basically, to achieve an intuitive and high-quality
deformation, the following criteria on the weights are demanded:
smoothness, non-negativity, partition-of-unity, locality/sparsity,
and no-local-maxima (see the analysis given in
\cite{jacobson2011bounded}).

The recent advancement of technology focuses on computing weights
of blending on a discrete form of domain (i.e., meshes are
employed to determine piecewise linear fields of weights). Weights
are computed on the mesh nodes via minimizing some discrete
differential energies (e.g., biharmonic, triharmonic and
quatraharmonic used in \cite{jacobson2012smooth}). After
incorporating the hard constraints according to the criteria on
weights, the weights are determined on mesh nodes with the help of
non-linear optimization. However, this is time-consuming. As a
result, the insertion of new handles cannot be realized in
real-time as new routines of non-linear optimization need to be
taken. Moreover, the determined weights are mesh-dependent. For a
symmetric shape to be deformed that is asymmetrically meshed, the
computed weights for a handle located at the symmetric positions
can
rarely be symmetric. %(see Fig.\ref{figChineseCharacter} for an example).
For poorly meshed computational domains, the artificial distortion
caused by the elements of poor shape is more serious (as
illustrated in Fig.\ref{fig:teaser}). Although the artifacts can
be reduced by increasing the density of meshes, this will further
slow down the computation. Ideally, the distribution of weights
should only be affected by the shape to be deformed and the
locations of handles, which indicates mesh-independence. Existing
mesh-independent approaches in literature for handle-driven
deformation (e.g.,
\cite{Singh1998Wires,Milliron2002Warps,vonFunck2006VectorField,Sumner2007EDS})
can only satisfy subsets of the demanded properties on weights.
This motivates our work on investigating a new meshfree method to
determine weights for shape deformation.

In this paper, we formulate the evaluation of weights in a
closed-form so that the deformation framework based on this gains
the benefit of flexibility -- i.e., the response of inserting new
handles is real-time. Specifically, the time cost of inserting a
new handle is linear to the number of samples used to represent
the domain of computation. The basis function formulated in this
approach can guarantee the properties of smoothness,
non-negativity, partition-of-unity, locality/sparsity, and
no-local-maxima, all of which are necessary to ensure a
deformation of high-quality.

The main results of our work are as follows:
\begin{itemize}
\item We present a meshfree method to determine linear blending
weights with $C^2$-continuity for real-time deformation. The
weights are formulated in a closed-form of basis functions
centered at the handles (details are given in Section
\ref{subsecFormulation}). After decomposing the region to be
deformed by the Voronoi diagram of handles, aforementioned
criteria of shape deformation are all ensured (see the analysis in
Section \ref{subsecAnalysis}).

\item A virtual handle insertion algorithm is proposed in Section
\ref{secHandle} to guarantee the locality and sparsity of
weighting so that a deformation interpolates the transformations
defined on handles. The virtual handles are added to let the
supporting region of the basis function defined on a handle not
cover the origins of any other handles (see the algorithm in
Section \ref{subsecInsertAlg}).

\item After constructing the Voronoi diagram of all handles
(including user-input and virtual ones), its dual-graph gives a
connectivity of the handles. We compute harmonic fields on the
graph to determine the transformations of virtual handles
according to the transformations specified on the user-input
handles (see Section \ref{subsecTransfOnVirtualHandle}). It is
found that the transformations determined in this way lead to a
shape-aware deformation following the intention of user input.
\end{itemize}
With the help of a discrete implementation on point samples
introduced in Section \ref{secImplementation}, an efficient and
effective meshfree approach has been developed for handle-driven
shape deformation. 2D/3D experimental results are shown in Section
\ref{secResult} to demonstrate the performance of our approach.

%%%%%%%%%%%%%%%%%%%%%%%%%%%%%%%%%%%%%%%%%%%%%%%%%%%%%%%%%%%%%%%%%%%%%%%%%%
%%%%%%%%%%%%%%%%%%%%%%%%%%%%%%%%%%%%%%%%%%%%%%%%%%%%%%%%%%%%%%%%%%%%%%%%%%
\section{Related Work}\label{secReview}
Shape deformation is an important research area in image
manipulation and geometric modeling. There are a large amount of
existing approaches in literature. The purpose of this section is
not for a comprehensive review. We only focus on discussing the
handle-driven deformation approaches.

Mesh-based techniques for discrete geometry modeling and
processing have been widely explored in the past decade. Typical
approaches including variational surface deformation
\cite{Botsch2004IFR}, Poisson deformation \cite{Yu2004PoissonDef},
Laplacian editing \cite{Sorkine2004Laplacian} and other linear
variational surface deformation approaches (see also the survey in
\cite{Botsch2008LinearDefSurvey}). Volumetric information and
rigidity are also incorporated to enhance the shape-preservation
in
\cite{Igarashi2005ARAP,Botsch2006PriMo,Botsch2007RigidCells,Sorkine2007SurfARAP}.
One common drawback of these approaches is that the positions of
vertices on a model need to be determined by solving a system of
linear equations after every update of handles, which becomes a
bottleneck of computation. A recent development in
\cite{jacobson2011bounded,Jacobson2011SA,jacobson2012smooth}
transfers the workload from online optimization to offline.
Specifically, the weights corresponding to handles are computed on
every vertex of a model before manipulating the handles (similar
to \cite{Zayer2005Harmonic}). The deformed shape is then evaluated
by linear blending of transformations defined on handles. In
\cite{Sumner2007EDS}, the handles are elements of a simplified
mesh. Although this strategy is more efficient than the
deformation methods based on online optimization, they still
cannot avoid solving large linear systems, which slows down the
response of deformation after inserting new handles. Moreover, the
results of deformation are also suffered from the artificial
distortions caused by the problems of meshes (e.g., too coarse
meshes for a fine deformation, a mesh with `needle' and `cap'
triangles, and the problem of symmetry). Our meshfree approach
solves these problems by providing closed-form formulas to
generate weights preserving all the demanded properties for
producing deformations with high quality in real-time.
%Vector field based shape deformation is defined in
%\cite{vonFunck2006VectorField} to construct a time-dependent
%divergence-free 3D  vector field to drive the
%deformation with $C^1$-continuity. However, it is not shape-aware.

Another thread of researches for deformation focuses on
mesh-independent approaches. Different handles are employed for
shape manipulation. Points are used in
\cite{Yoshizawa2002Def,Schaefer2006MLS}, and curves are employed
as handles in \cite{Lazarus1994AxialDef,Singh1998Wires}.
Grid-based deformation techniques in
\cite{Sederberg1986FFD,Lee1995IMU} conduct the
bivariate/trivariate cubic splines to realize deformations with
$C^2$-continuity. Users are allowed to move control points of the
spline surfaces/solids to modify the embedded shapes, where the
editing is indirect. Some approaches have been developed to extend
this approach to provide the ability of direct editing
(ref.~\cite{Hsu1992DirectFFD,Hu2001DirectFFD}). However, the
computational domain is still limited to a simple topology (i.e.,
genus zero). An improvement of the grid-based techniques is
introduced by Beier and Neely~\shortcite{Beier1992FIM} to allow
handles in the form of line segments by using the Shepard's
interpolation \cite{Shepard1968}. Cage-based deformation (e.g.,
\cite{Joshi2007HarmonicCoordinates,Ben-Chen2009VariationalHarmonic})
can be considered as a further generalization of grid-based
deformation, where weights can be found by a closed-form in terms
of the handles in
\cite{Ju2005MeanValue,Lipman2008GreenCoordinates}. However, the
construction of cages is usually not automatic and the
manipulation on cages instead of a model itself is indirect.

\textit{Moving least square} (MLS) strategy is employed in
\cite{Schaefer2006MLS} for interpolating the similarity/rigid
deformation at handle points. A closed-form solution is provided
in their approach to determine the transformation matrix on every
point in a MLS manner. The transformations in the whole domain
need to be computed when any handle is moved. In other words, the
deformation is globally affected by all handles -- lack of
sparsity. Different from this MLS approach, our approach belongs
to the category of linear blending based deformation. When the
property of sparsity is preserved on the weights, the deformation
at a point is only affected by the nearby handles that is easier
to be predicted by end-users. Moreover, the deformation determined
by our approach is resolution independent, which is very important
for image manipulation.

The work of generating weights for linear blending also relates to
the research of scattered data interpolation, where \textit{radial
basis functions} (RBF) are widely used (e.g.,
\cite{Floater1996RBF,Botsch2005RBF}). In \cite{Botsch2005RBF}, the
deformation is governed by global RBFs that lead to a dense linear
system to be solved. The weights determined by the dense (or
global) data interpolation approaches lack of sparsity. Therefore,
every point in the domain is changed when any handle is updated
even if it is far away. Although the \textit{compactly supported
radial basis functions} (CSRBF) can help on introducing the
sparsity (ref. \cite{Floater1996RBF}), it does not provide
closed-form formulas as our approach.

%%%%%%%%%%%%%%%%%%%%%%%%%%%%%%%%%%%%%%%%%%%%%%%%%%%%%%%%%%%%%%%%%%%%%%%%%%
%%%%%%%%%%%%%%%%%%%%%%%%%%%%%%%%%%%%%%%%%%%%%%%%%%%%%%%%%%%%%%%%%%%%%%%%%%
\section{Meshfree Weighting}\label{secWeighting}
Following the linear blending formulation, the new position of a
point $\mathbf{p} \in \Omega$ is determined by the transformations
$\mathbf{T}_i$ defined on handles $H_i$ as\footnote{$\mathbf{T}_i$
is a homogenous matrix and $\mathbf{p}$ is represented by
homogeneous coordinate.}
\begin{equation}\label{eqDeformation}
\mathbf{p}'=\sum_{i=1}^m w_i(\mathbf{p})\mathbf{T}_i \mathbf{p}
\end{equation}
with $w_i(\cdot)$ being the scalar field of weights to be
determined. The origin of a handle $H_i$ is denoted by
$\mathbf{h}_i$. This linear blending based deformation is fast and
easy-to-implement. However, carelessly assigned weights can lead
to visible artifacts in results. Basically, a deformation with
high quality must have the following properties:
\begin{itemize}
\item \textbf{Smoothness:} The scalar field of weights must be
smooth to avoid visual artifact (discontinuity) in both 2D and 3D
deformations. %Many prior works (see \cite{Botsch2008LinearDefSurvey}) devote their attention to improve the smoothness of deformation.
We use compactly supported B\'{e}zier basis functions in our
formulation, which lead to a weight field with $C^2$-continuity.

\item \textbf{Interpolation:} The final transformation determined
by the linear blending must interpolate the transformations at the
handles. Specifically, the weight on a handle $H_i$ is
\textit{one} at its origin while basis functions centered at other
handles give \textit{zero} at this point.
%That is,
%\begin{center}
%$w_i(\mathbf{h}_i)=1$ and $w_{j, \forall j \neq i}(\mathbf{h}_i)=0$.
%\end{center}
%This property is also demanded at the vicinity of handles.
%Generally, we may wish the deformation of a point is only
%determined by a handle $H_i$ when the point's distances to other
%handles are much farther than its distance to $H_i$.
This is
guaranteed by the locality and the sparsity in our formulation.

\item \textbf{Consistency:} When applying the same transformation
$\mathbf{T}$ on all handles, all points in $\Omega$ must be
consistently transformed by $\mathbf{T}$. This is enforced by the
partition-of-unity property in our formulation.
%\begin{center}
%$\forall \mathbf{p} \in \Omega$, $\sum_{i=1}^m w_i(\mathbf{p}) \equiv 1$.
%\end{center}
Another consistency requirement is about direction. The region
influenced by a handle should not change in the inverse direction
of the transformation assigned on the handle. We ensure this by
the property of non-negativity.

\item \textbf{Shape-awareness:} This is a property more or less
subjective. Basically, the intrinsic requirement on
shape-awareness is to have deformations like stretching, bending
and twisting an elastic solid, where the handles serve as pins.
%When realizing the shape-aware deformation by linear blending, the
%weight $w_i(\cdot)$ according to a handle $H_i$ is expected to be
%\emph{zero} for points outside the region influenced by $H_i$ and
%be constant \emph{one} for points \emph{only} influenced by $H_i$.
%This is actually the property of interpolation analyzed above. For
%the point in a region that is influenced by multiple handles, we
%wish to have monotonic decay of a handle's influence -- that is
%the weights of different handles are inverse proportional to the
%distances to the handles.
In our formulation, this is preserved by
1) having non-positive first derivative of basis functions and 2)
letting all basis functions have similar support sizes.
No-local-maxima on weights will prevent generating singularity
(e.g., a point moves faster than all its neighbors) during
deformation.
\end{itemize}
Our formulation below leads to $C^2$-continuous weights preserving
all these properties in deformations.

\begin{figure}
\includegraphics[width=\linewidth]{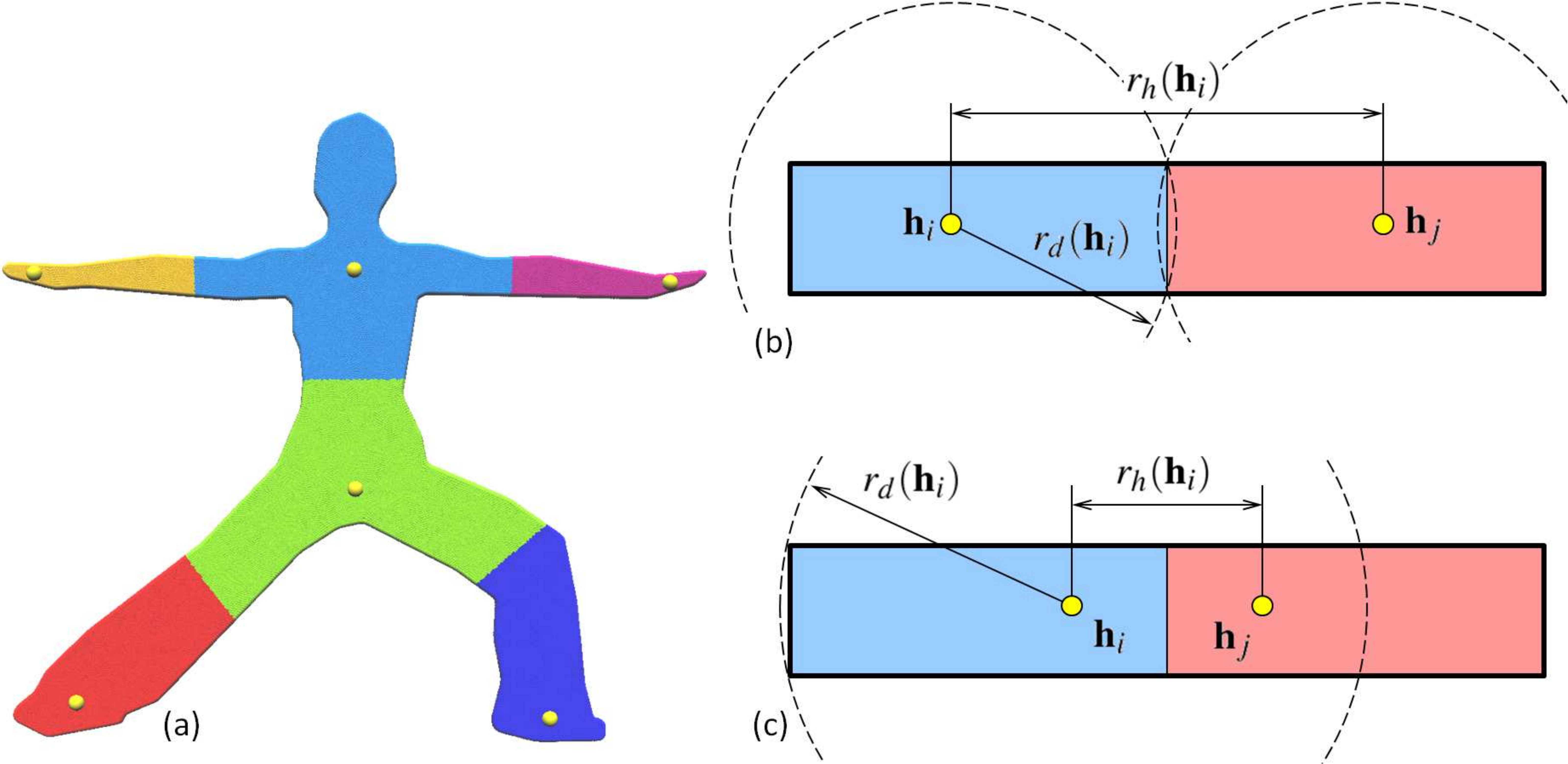}
\caption{Voronoi diagram based method to determine the size of
local support. (a) The Voronoi diagram of handles can decompose
$\Omega$ into smaller pieces. (b) The illustration of
$r_h(\mathbf{h}_i)$ and $r_d(\mathbf{h}_i)$ in the Voronoi
diagram. (c) Very close handles can lead to $r_h(\mathbf{h}_i) <
r_d(\mathbf{h}_i)$.}\label{figVoronoiDiagram}
\end{figure}

\subsection{Formulation}\label{subsecFormulation}
Each handle $H_i$ is equipped with a compactly supported basis
function with support size $r_i$ as $\phi_i(d(\mathbf{p},
\mathbf{h}_i)/r_i)$, where $\mathbf{h}_i$ is the location of $H_i$
and $d(\cdot,\cdot)$ returns the intrinsic-distance (see Appendix
A for the definition) between two points inside $\Omega$. The
scalar field of the weights for $H_i$ is then defined as
\begin{equation}\label{eqWeight}
w_i(\mathbf{p})=\frac{\phi_i(d(\mathbf{p}, \mathbf{h}_i)/r_i)}
{\sum_{j=1}^m \phi_j(d(\mathbf{p}, \mathbf{h}_j)/r_j)},
\end{equation}
which enforces the partition-of-unity.

To be shape-aware and interpolate handles, $\phi_i(\cdot)$ is
chosen as a monotonically decreasing function with $\phi_i(0)=1$
and $\phi_i(t)=0$ ($\forall t \geq 1$). A quintic polynomial is
employed for the function $\phi_i(t)$ so that the constraints for
$C^1$ and $C^2$-continuity at the boundary of the supporting
regions can be satisfied. Specifically, we need
\begin{equation}\label{eqConstraints}
\phi_i'(0)=\phi_i'(1)=\phi_i''(0)=\phi_i''(1)=0.
\end{equation}
To ease the evaluation and analysis, each $\phi_i(t)$ is
represented as the $y$-component (i.e.,
$\phi_i(t)=\mathbf{b}^y(t)$, $t\in [0,1]$) of a 2D B\'{e}zier
curve with degree-$n$ ($n \geq 5$)
\begin{equation}\label{eqBezier}
\mathbf{b}(t)=\sum_{i=0}^n \mathbf{b}_i B_{i,n}(t),
\end{equation}
where $B_{i,n}(t)$ are the Bernstein polynomials. From the
property of B\'{e}zier curves (ref. \cite{Farin2001CAGD}), we know
that $x=t$ when $\mathbf{b}_i^x=i/n$. Letting
$\mathbf{b}_{0,1,2}^y \equiv 1$ and $\mathbf{b}_{n,n-1,n-2}^y
\equiv 0$ can satisfy these constraints at the endpoints (see
Appendix B for more details). For the rest control points, we can
simply assign them as $0.5$ or align them along the line
$\mathbf{b}_2 \mathbf{b}_{n-2}$ uniformly.

When the intrinsic-distance is used to generate the input
parameter $t$ for the basis functions, linear blending based
deformations driven by these basis functions behave in a
shape-aware manner. Now the problem left is how to determine the
support size $r_i$ of each basis function. As a basic requirement
of handle-driven deformation based on linear blending, every point
$\mathbf{p} \in \Omega$ should be influenced by at least one
handle. To be shape-aware, a point $\mathbf{p}$ should be mostly
affected by its closest handle in $\Omega$. Voronoi diagram sited
at the origins of handles $\{ \mathbf{h}_i \}$ provides an
intrinsic decomposition of $\Omega$ according to these
observations (see Fig.\ref{figVoronoiDiagram}(a)), where the
intrinsic-distance in $\Omega$ is used as the metric for
generating the Voronoi diagram. We denote the cell that
corresponds to $\mathbf{h}_i$ by $\mathcal{V}(\mathbf{h}_i)$. Two
metrics according to a handle $H_i$ can be defined as follows (see
Fig.\ref{figVoronoiDiagram}(b) for an illustration):
\begin{itemize}
\item The size of a Voronoi cell: $r_d(\mathbf{h}_i) =
\sup_{\mathbf{q} \in \mathcal{V}(\mathbf{h}_i)}
d(\mathbf{q},\mathbf{h}_i)$;

\item The separation to other sites: $r_h(\mathbf{h}_i) =
\inf_{\mathbf{h}_{j \; (j \neq i)}} d(\mathbf{h}_i,\mathbf{h}_j)$.
\end{itemize}
%\begin{itemize}
%\item The maximal distance to site in the cell of Voronoi diagram
%\begin{displaymath}
%r_d(\mathbf{h}_i) = \sup_{\mathbf{q} \in \mathcal{V}(\mathbf{h}_i)} d(\mathbf{q},\mathbf{h}_i).
%\end{displaymath}
%\item The minimal distance to other sites:
%\begin{displaymath}
%r_h(\mathbf{h}_i) = \inf_{\mathbf{h}_{j \; (j \neq i)}} d(\mathbf{h}_i,\mathbf{h}_j).
%\end{displaymath}
%\end{itemize}
To let the basis function $\phi_i(t)$ centered at $H_i$ cover all
points in $\mathcal{V}(\mathbf{h}_i)$ and to ensure the handle
interpolation property, it should have
\begin{equation}\label{eqSupportSize}
r_d(\mathbf{h}_i)< r_i \leq r_h(\mathbf{h}_i).
\end{equation}
The support size can be $r_i = (1-\alpha) r_d(\mathbf{h}_i) +
\alpha r_h(\mathbf{h}_i)$ with $\alpha \in (0,1]$ being specified
by users as a shape factor. For most of the examples in this
paper, $\alpha=1$ is used. It is possible to have two handles too
close to each other so that $r_h(\mathbf{h}_i) <
r_d(\mathbf{h}_i)$ (see Fig.\ref{figVoronoiDiagram}(c) for an
example). For solving such cases, we will use the virtual handle
insertion algorithm (presented in Section \ref{secHandle}).

\subsection{Analysis and Discussion}\label{subsecAnalysis}
We analyze the advantages of our formulation for the handle-driven
deformation based on linear blending.

\textbf{Non-negativity:} $\phi_i(t) \geq 0$ so that $\forall
\mathbf{p} \in \Omega, w_i(\mathbf{p}) \geq 0$. Moreover, when
$r_i > r_d(\mathbf{h}_i)$ is ensured for all handles, every point
in $\Omega$ should be covered by at least one handle's support. In
other words, $\sum_{j=1}^m \phi_j(\cdot) \neq 0$.

\textbf{Partition-of-unity:} This has been enforced by the
formulation in Eq.(\ref{eqWeight}). That is,
\begin{displaymath}
\sum_{i=1}^m w_i(\mathbf{p})=\sum_{i=1}^m
\frac{\phi_i(d(\mathbf{p}, \mathbf{h}_i)/r_i)} {\sum_{j=1}^m
\phi_j(d(\mathbf{p}, \mathbf{h}_j)/r_j)} \equiv 1.
\end{displaymath}

\textbf{Locality/Sparsity:} This is preserved by $\forall t\geq 1,
\phi_i(t)\equiv 0$ and the condition given in
Eq.(\ref{eqSupportSize}). The transformation at a point coincident
with a handle is only determined by the handle itself. $\forall i
\neq j, \phi_j(\mathbf{h}_i) \equiv 0$.

\textbf{Smoothness:} $C^2$-continuity is preserved on the weights
determined by Eq.(\ref{eqWeight}). First of all, the basis
function $\phi_i(t)=\mathbf{b}^y(t)$ is $C^n$-continuous for $t
\in (0,1)$ when $\mathbf{b}^y(t)$ is defined as a B\'{e}zier curve
in Eq.(\ref{eqBezier}) with $n \geq 5$. Therefore,
$w_i(\mathbf{p})$ is also $C^n$-continuous when $
\phi_{j}(d(\mathbf{p}, \mathbf{h}_j)/r_j) \neq 0$ for any other $j
\neq i$. In the region that is only covered by the support of
$H_i$, $w_i \equiv 1$. Similarly, it is also a constant function
($w_i \equiv 0$) in the region outside the support of $H_i$. By
Eq.(\ref{eqConstraints}), it is not difficult to prove the
$C^2$-continuity at the following two cases: \\i) $d(\mathbf{p},
\mathbf{h}_j)<r_j$ and $d(\mathbf{p}, \mathbf{h}_i) = r_i$, \\ ii)
$d(\mathbf{p},\mathbf{h}_j) = r_j$ and $d(\mathbf{p}, \mathbf{h}_i) < r_i$, \\
where both the first and second derivatives are zero.

\textbf{No-local-maxima:} The global maxima of a weight $w_i$ only
happens at the origin of handle $H_i$ and the regions only covered
by the support of $H_i$. Besides, we also observe the phenomenon
of no-local-maxima in all our experimental tests.

\textbf{Closed-form:} The weights $\{w_i (\mathbf{p}) \}$ at any
point $\mathbf{p} \in \Omega$ are evaluated in a closed-form
(i.e., by Eq.(\ref{eqWeight})). This guarantees the flexibility of
inserting new handles during the deformation in real-time.

\textbf{Meshfree:} As the evaluation of basis functions to
determine the weights is only related to the intrinsic-distance
from points to the origin of handles, the solution is independent
of mesh quality and resolution. In the mesh-dependent solutions,
elements with poor shape, which can occur after a drastic
deformation step, must be optimized. Remeshing leads to another
round of weights computation that could be time-consuming.

In short, our method preserves all the merits of prior methods for
linear blending based deformation (e.g.,
\cite{jacobson2011bounded,Jacobson2011SA,jacobson2012smooth})
while introducing new benefits of flexibility and efficiency.

\begin{figure}
\includegraphics[width=\linewidth]{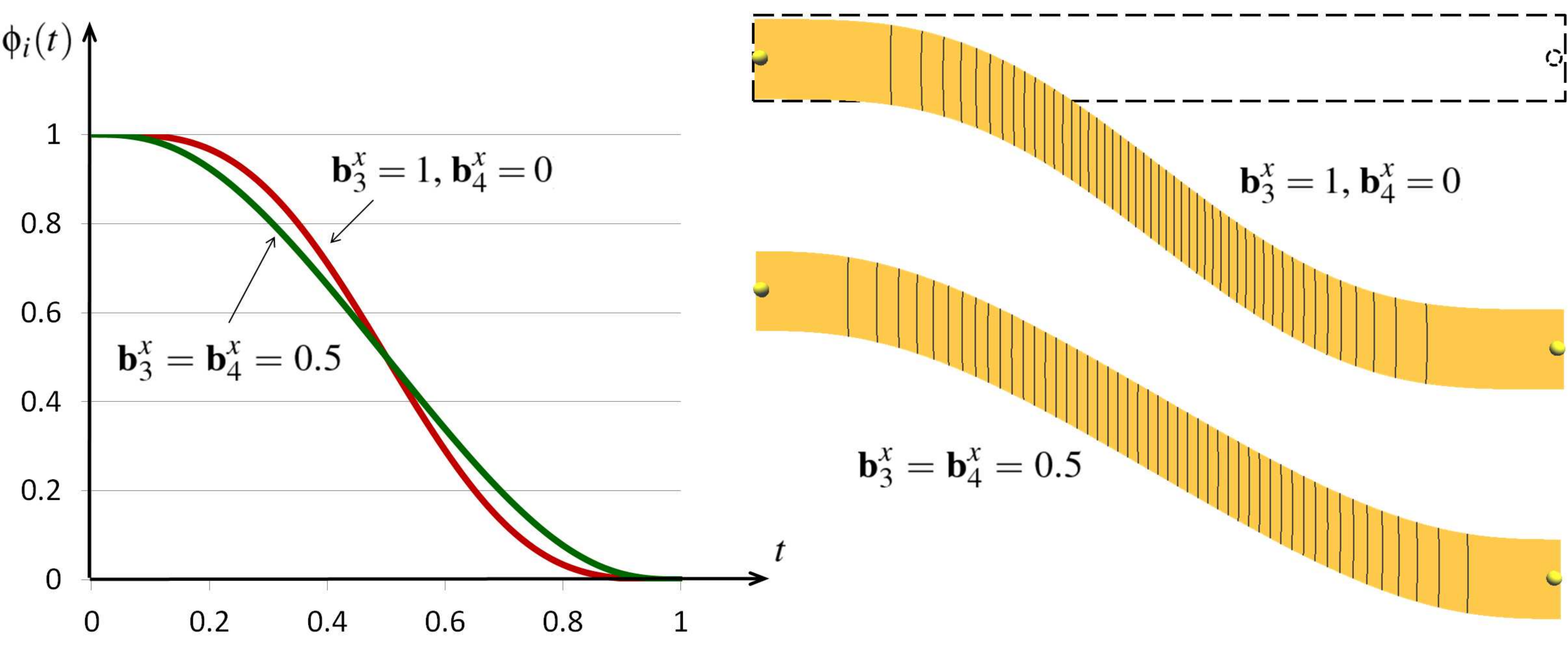}
\caption{By using different basis functions in our formulation,
different distributions of weights can be obtained which lead to
the change of deformation behaviors. Isocurves for the weight
field of the right handle are also shown in black lines on the
deformed bars.}\label{figBasisFunc}
\end{figure}

Besides the flexibility of inserting new handles during the
deformation, we also provide users a method to change the behavior
of handles by adjusting the shape of basis functions (i.e.,
$\phi_i(t)$). For example, as shown in Fig.\ref{figBasisFunc}, for
the basis function $\phi_i(t)$ built by a septic B\'{e}zier curve
($n=7$), we can assign different values to $\mathbf{b}^x_3$ and
$\mathbf{b}^x_4$ to obtain different shapes for $\phi_i(t)$ to
have different deformation behaviors. Basically, a `flat' basis
function (e.g., $\mathbf{b}^x_3=\mathbf{b}^x_4=0.5$) results in a
deformation simulating hard materials while a more curved basis
function (e.g., $\mathbf{b}^x_3=1$, $\mathbf{b}^x_4=0$) makes the
deformation soft. When using polynomials in higher orders, we have
more degree-of-freedoms to change the shape of basis function.
However, according to our experiments, septic polynomials are good
enough in most of the cases.
%When using polynomials in different orders, the basis functions
%have different shapes therefore lead to different deformation
%behaviors (see Fig.\ref{} for examples).

The formulation of meshfree weighting also has some limitations.
First, the interpolation property cannot be preserved when the
distance between two handles are too close while the regions to be
covered by either handle are large. Specifically, the
interpolation of handles becomes an approximation when
$r_d(\mathbf{h}_i) < r_h(\mathbf{h}_i)$ in
Eq.(\ref{eqSupportSize}) can \textit{NOT} be satisfied. Second,
for the region that is only covered by one handle, the
transformation is consistent with the handle. Then, the
deformation presented in this region is not shape-aware -- i.e.,
the influence of handles does not decay while increasing the
distance to the handle's center. Both the problems will be solved
by applying the virtual handle insertion approach presented in the
following section.

%%%%%%%%%%%%%%%%%%%%%%%%%%%%%%%%%%%%%%%%%%%%%%%%%%%%%%%%%%%%%%%%%%%%%%%%%%
%%%%%%%%%%%%%%%%%%%%%%%%%%%%%%%%%%%%%%%%%%%%%%%%%%%%%%%%%%%%%%%%%%%%%%%%%%
\section{Virtual Handle Insertion}\label{secHandle}
A handle insertion algorithm is developed to enrich our meshfree
weighting framework in the aspects of guaranteeing the handle
interpolation property and improving the shape-awareness of
deformation.

\subsection{Insertion algorithm}\label{subsecInsertAlg}
When $r_d(\mathbf{h}_i) > r_h(\mathbf{h}_i)$, %according to the definition of $r_d(\mathbf{h}_i)$ and $r_h(\mathbf{h}_i)$
we know
that there are points in the voronoi cell
$\mathcal{V}(\mathbf{h}_i)$ whose distances to $\mathbf{h}_i$ are
larger than the minimal distance from $\mathbf{h}_i$ to other
handles.

\textbf{Proposition 1} \hspace{5pt} When $r_d(\mathbf{h}_i) >
r_h(\mathbf{h}_i)$, inserting new sites at the points
$\mathbf{h}_d \in \mathcal{V}(\mathbf{h}_i)$ with $d(\mathbf{h}_d
, \mathbf{h}_i) = r_d(\mathbf{h}_i)$ can reduce
$r_d(\mathbf{h}_i)$ while keeping $r_h(\mathbf{h}_i)$ unchanged.
\begin{proof}
First of all, the value of $r_h(\mathbf{h}_i)$ is not affected.
When $\mathbf{h}_d$ is the only point in
$\mathcal{V}(\mathbf{h}_i)$ with $d(\mathbf{h}_d , \mathbf{h}_i) =
r_d(\mathbf{h}_i)$, it is obvious $\exists \mathbf{q} \in
\mathcal{V}(\mathbf{h}_i)$ with $d(\mathbf{q},\mathbf{h}_d) <
d(\mathbf{q},\mathbf{h}_i)$. Define $\mathcal{S}(\mathbf{h}_d)=\{
\mathbf{q} \in \mathcal{V}(\mathbf{h}_i) \; | \;
d(\mathbf{q},\mathbf{h}_d) < d(\mathbf{q},\mathbf{h}_i) \}$. After
inserting a new site at $\mathbf{h}_d$, the points in
$\mathcal{S}(\mathbf{h}_d)$ become the member of
$\mathcal{V}(\mathbf{h}_d)$. When all points with $d(\mathbf{h}_d
, \mathbf{h}_i) = r_d(\mathbf{h}_i)$ have been assigned to other
voronoi cells, the value of $r_d(\mathbf{h}_i)$ reduces. On the
other aspect, the distances from the newly inserted points to
$\mathbf{h}_i$ are $r_d(\mathbf{h}_i)$ which is greater than
$r_h(\mathbf{h}_i)$.
\end{proof}

Based on this proposition, we develop a greedy algorithm for
handle insertion. Define $\mathcal{H}$ as the set of handles and
$\delta(\cdot)=r_d(\cdot)-r_h(\cdot)$. When $\exists \mathbf{h}_i
\in \mathcal{H}$ with $\delta(\mathbf{h}_i)>0$, new handles are
inserted to resolve this problem by reducing $\max_{\mathbf{h}_i
\in \mathcal{H}} \{ \delta(\mathbf{h}_i) / r_d(\mathbf{h}_i) \}$.
The pseudo-code is described as \textbf{Algorithm} \textit{Virtual
Handle Insertion}.
\begin{algorithm}[h]
\caption{\textit{Virtual Handle Insertion}}

%\LinesNumbered
%\linesnumbered

%\SetAlgoNoLine

\KwIn{the set $\mathcal{H}$ of real handles}

\KwOut{the expanded set $\mathcal{H}$ with virtual handles}

\While{$\exists \mathbf{h}_i \in \mathcal{H}$,
$\delta(\mathbf{h}_i)>0$}{

Find the handle $\mathbf{h}_m = \arg \max_{\mathbf{h}_i \in
\mathcal{H}} \delta(\mathbf{h}_i) / r_d(\mathbf{h}_i)$;

Find a point $\mathbf{p} \in \mathcal{V}(\mathbf{h}_m)$ with
$d(\mathbf{p}, \mathbf{h}_m) = r_d(\mathbf{h}_m)$;

Insert a new handle located at $\mathbf{p}$ into $\mathcal{H}$;

Update the values of $r_d(\cdot)$ and $r_h(\cdot)$ on all handles;

}

\Return $\mathcal{H}$;

\end{algorithm}

\begin{figure}\centering
\includegraphics[width=\linewidth]{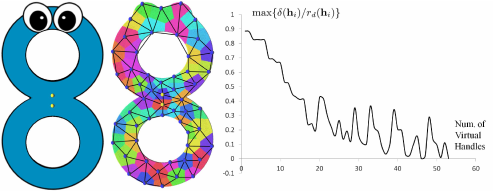}
\caption{When two handles are too close to each other (see left),
the condition for interpolation (i.e., $r_d(\cdot) < r_h(\cdot)$)
can only be satisfied after inserting virtual handles. (Middle)
The newly inserted virtual handles (in blue) tessellate the
deformation domain into voronoi cells whose areas are similar to
neighboring cells. The Delaunay graph, $\mathcal{DG}(\mathcal{H}
\cup \mathcal{H}^v)$, of the voronoi diagram is also shown -- see
the network linking the handles. (Right) The score,
$\max\{\delta(\mathbf{h}_i) / r_d(\mathbf{h}_i)\}$, of our Virtual
Handle Insertion algorithm drops while inserting virtual
handles.}\label{figVirtualHandleTori}
\end{figure}

%\begin{enumerate}
%\item Ths.
%
%\item The new handle $H_d$ is inserted into $\mathcal{H}$;
%
%\item Iterating the handle insertion until there is no handle with $r_d(\cdot) > r_h(\cdot)$.
%\end{enumerate}
%As the result of this algorithm, a new set of handles can be obtained. Denote this new set as $\mathcal{H}^*$.

\textit{Remarks.} \hspace{5pt} From \textit{Proposition} 1, we
know that inserting new handles in a voronoi cell
$\mathcal{V}(\mathbf{h}_i)$ with $\delta(\mathbf{h}_i) > 0$ can
reduce the value of $\delta(\mathbf{h}_i)$. However, inserting a
new site $\mathbf{h}_d \in \mathcal{V}(\mathbf{h}_i)$ can also
affect the other handles (i.e., $H_j$ with $j \neq i$). In extreme
cases, the original $\delta(\mathbf{h}_j)<0$ could be turned into
$\delta(\mathbf{h}_j)>0$. Then, new handles need to be added into
$\mathcal{V}(\mathbf{h}_j)$.

Our virtual handle insertion algorithm can be considered as a
variant of the farthest point sampling algorithm, which tends to
tessellate a domain into a voronoi diagram with neighboring
voronoi cells having similar sizes. The condition of
$r_d(\cdot)<r_h(\cdot)$ is satisfied on all handles when this is
the case. Our experimental tests also follow this observation (see
Fig.\ref{figVirtualHandleTori} for an example).

\begin{figure}[t]\centering
\includegraphics[width=\linewidth]{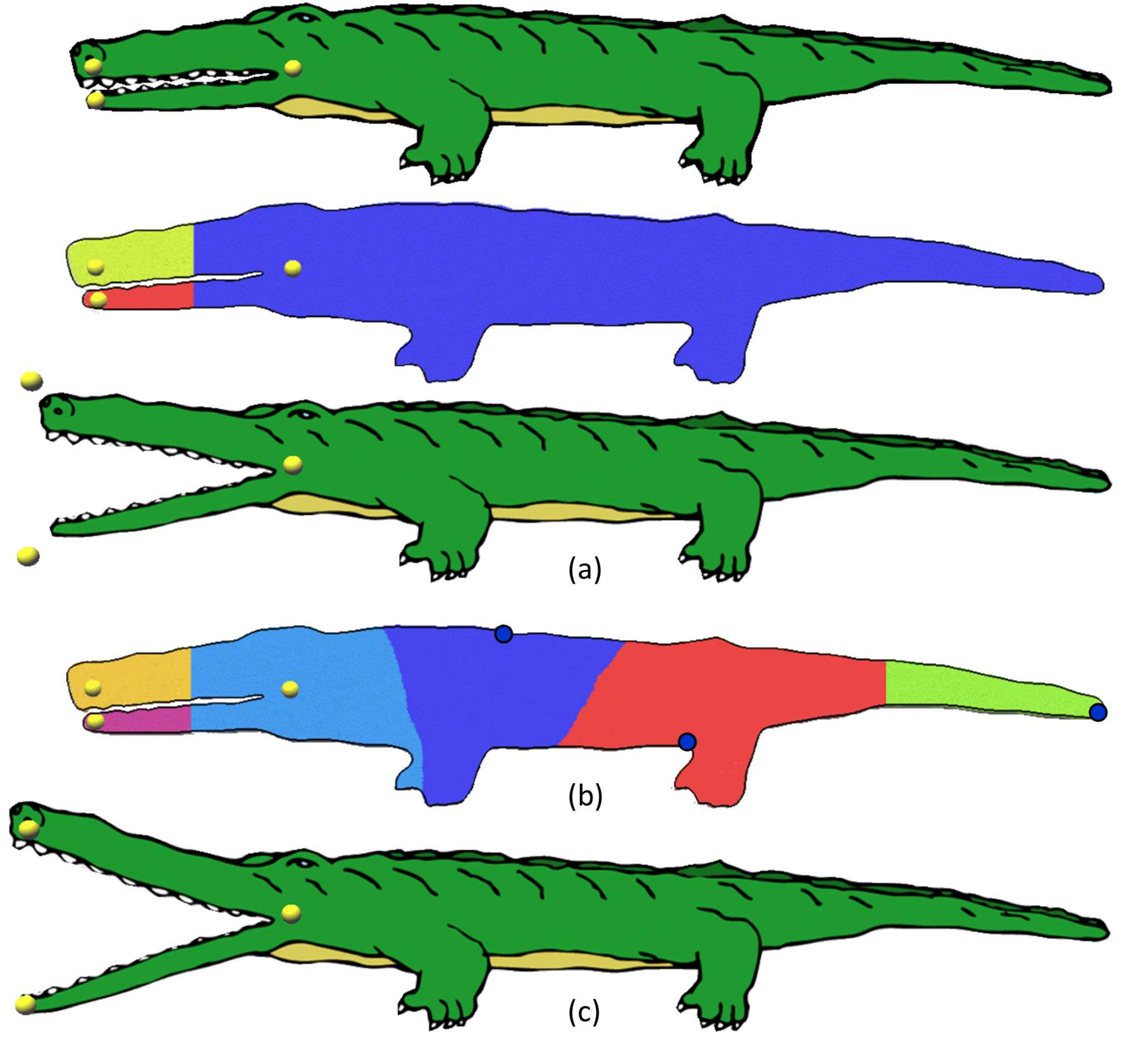}
\caption{A handle covering a large region can affect the
interpolation on its nearby handles. (a) For the handle at the
right, its voronoi cell covers all the right part of the alligator
-- this leads to a value of $r_d(\cdot)$ that is much larger than
$r_h(\cdot)$. In this case, transformations at the left two
handles cannot be interpolated. (b) Virtual handles (in blue
color) are added to resolve the problem by the \textit{insertion}
algorithm. As a result, the domain to be deformed has been
decomposed into smaller voronoi
cells with handles (real and virtual) as sites. %It is easy to observe that the neighboring voronoi cells have similar area.
(c) The deformation result is driven by both the real and the
virtual handles, where the transformations at real handles are
interpolated.}\label{figVirtualHandle}
\end{figure}

\subsection{Transformation on Virtual Handles}\label{subsecTransfOnVirtualHandle}
\begin{figure}\centering
\includegraphics[width=\linewidth]{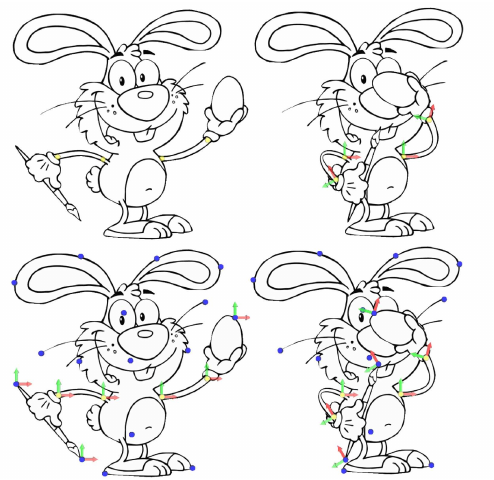}
\caption{The deformation of a rabbit is drive by four real handles
(see the yellow dots and the frames shown in the top row). The
result of deformation is determined with the help of virtual
handles (shown in blue dots). The transformations at handles (both
real and virtual ones) are illustrated by
frames.}\label{figRabbit}
\end{figure}

A left problem is how to determine the transformation on virtual
handles according to the user-specified transformations on real
handles. Denote the set of real handles as $\mathcal{H}$ and the
set of virtual handles as $\mathcal{H}^v$. As aforementioned, the
handles of $\mathcal{H} \cup \mathcal{H}^v$ have partitioned the
given domain $\Omega$ into a voronoi diagram
$\mathit{Vor}(\mathcal{H} \cup \mathcal{H}^v)$. A dual graph of
$\mathit{Vor}(\mathcal{H} \cup \mathcal{H}^v)$ can be constructed
by 1) using the sites of every voronoi cell as nodes and 2)
linking the sites of every two neighboring voronoi cells by a
straight line, which is a Delaunay graph \cite{Berg2008CompGeo}.
We denote the Delaunay graph by $\mathcal{DG}(\mathcal{H} \cup
\mathcal{H}^v)$ and also use symbol $H$ to represent nodes in
$\mathcal{DG}$ since each node is in fact a handle (real or
virtual). The transformations of handles in $\mathcal{H}^v$ are
determined with the help of the Delaunay graph as follows.
\begin{itemize}
\item For each handle $H_i$ in $\mathcal{H}$, a harmonic field
$\varpi_i(\cdot)$ is computed on $\mathcal{DG}$ to assign each
handle $H_g$ a field value $\varpi_i(H_g)$. Boundary conditions,
$\varpi_i(H_i)=1$ and $\varpi_i(H_{j \neq i})=0$, are given to
compute the harmonic field $\varpi_i(\cdot)$. If there are $m$
handles in $\mathcal{H}$, $m$ harmonic fields are determined on
$\mathcal{DG}$.

\item After converting the transformation $\mathbf{T}_i$ of each
handle into a rotation quaternion $\mathbf{q}_i$ and a translation
vector $\mathbf{t}_i$, the rotation and the translation on a
virtual handle $H_v \in \mathcal{H}^v$ can be determined by
\begin{equation}\label{eqTransBlendingOnVirtualHandle}
\left(%
\begin{array}{c}
  \mathbf{q}_v \\
  \mathbf{t}_v \\
\end{array}%
\right)=\frac{1}{\varpi_{sum}(H_v)} \sum_{H_i \in \mathcal{H}} \varpi_i(H_v)\left(%
\begin{array}{c}
  \mathbf{q}_i \\
  \mathbf{t}_i \\
\end{array}%
\right)
\end{equation}
with $\varpi_{sum}(\cdot) = \sum_{H_j \in \mathcal{H}}
\varpi_j(\cdot)$.

\item Finally, the quaternion and the translation determined on
each virtual handle are converted back into a transformation
matrix to be used in linear blending.
\end{itemize}
The transformation of virtual handles determined in this way
brings in the effect of shape-awareness during the deformation. As
illustrated in Figs.\ref{figVirtualHandle} and \ref{figRabbit},
the deformation of whole domain driven by the transformations on
handles (real and virtual) is very natural. The influence of a
real handle decays when the distance to it increases.

%%%%%%%%%%%%%%%%%%%%%%%%%%%%%%%%%%%%%%%%%%%%%%%%%%%%%%%%%%%%%%%%%%%%%%%%%%
%%%%%%%%%%%%%%%%%%%%%%%%%%%%%%%%%%%%%%%%%%%%%%%%%%%%%%%%%%%%%%%%%%%%%%%%%%
\section{Implementation Details}\label{secImplementation}
%This section discusses some implementation details of our
%approach, including the primitives of computation, the approximate
%evaluation and the extension to more general types of handles.
%
Similar to many other meshfree approaches, we sample the input
domain $\Omega$ to be deformed into a set of dense points
$\mathcal{P}$. By searching $k$-nearest-neighbors of each point, a
graph $\mathcal{G}(\mathcal{P})$ spanning $\Omega$ (in discrete
form) can be established by using points in $\mathcal{P}$ as nodes
and adding links between neighboring points. Note that user
specified handles should also be added into $\mathcal{P}$ to
construct the graph (i.e., $\mathcal{H} \subset \mathcal{P}$). The
intrinsic-distance from any point $\mathbf{q} \in \mathcal{P}$ to
a handle is approximated by the distance between $\mathbf{q}$ and
the handle on the graph, which can be computed efficiently with
the help of Dijkstra's algorithm. Also, the voronoi diagram
$\mathit{Vol}(\mathcal{H})$ can be obtained by the Dijkstra's
algorithm with multiple sources on $\mathcal{G}(\mathcal{P})$,
where each sample is assigned to a voronoi cell. As the primitives
used in the computation are points, the deformation approach can
be easily generalized from 2D images to 3D solids. More examples
can be found in the following section. To determine the weights on
a general point $\mathbf{p} \in \Omega$ that is not a sample in
$\mathcal{P}$, a linear blending based on reciprocal distance
weights \cite{Floater01MeshlessPara} is employed to obtain the
weight on $\mathbf{p}$ from its $k$-nearest-neighbors in
$\mathcal{P}$. There are more sophisticated parameterization
strategies in \cite{Floater01MeshlessPara}, which can also be
applied here. With the help of this meshless parameterization, we
can easily take an up-sampling step in the domain $\Omega$ when
the point set $\mathcal{P}$ becomes sparse when applying a drastic
deformation.

After using the virtual handle insertion algorithm to generate a
set of new handles, harmonic fields are computed on a dual graph
of $\mathit{Vol}(\mathcal{H})$ to determine the transformations on
virtual handles. By our boundary condition, all field values are
non-negative when uniform Laplacian is employed
\cite{Wardetzky2007Laplacian}. In other words, the coefficients
used in Eq.(\ref{eqTransBlendingOnVirtualHandle}) are
non-negative. Instead of solving a linear system to compute the
harmonic field, we initially assign the field values on all real
handles as one and the weights on all virtual handles are set as
zero. Then we apply Laplacian operators to update their field
values iteratively. The field values on virtual handles can be
efficiently obtained after tens of iterations.

\begin{figure}\centering
\includegraphics[width=\linewidth]{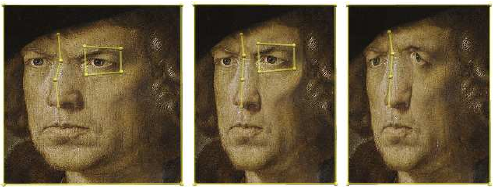}
\caption{A portrait is edited by segment handles. Salient feature
inside the closed loop of segment handles at the left eye is
preserved after the deformation.}\label{figPortrait}
\end{figure}

The point handles can be generalized to different types of handles
(e.g., line segments and polygons, etc.). Specifically, each
handle $H_g$ now becomes a set of points $\{\mathbf{h}_g\}$
instead of a single point while all these points are equipped with
the same transformation $\mathbf{T}_g$. The major change is the
method to evaluate the intrinsic-distance from a query point
$\mathbf{q}$ to handles (e.g., line segments), which is the
intrinsic-distance to $\mathbf{q}$'s closest sample point on the
handle. The rest of our approach will keep unchanged. Extreme case
occurs when two line-segment handles have a common endpoint so
that $r_h(\cdot)$ of these two handles becomes zero. There is no
way to satisfy the condition of $r_d(\cdot)<r_h(\cdot)$ for handle
interpolation. We therefore only approximate the transformations
specified on handles. Specifically, the basis function is changed
to a global Gaussian
\begin{equation}\label{eqBasisForApprox}
\phi_i(t)=e^{-(c_i t)^2}
\end{equation}
with $c_i$ being a constant to control the width of Gaussian. In
our implementation, letting $c_i$ be $\frac{1}{2} r_h(\cdot)$
works well in all tests. As some handles may have common
endpoints, $r_h(\cdot)$ is changed to the minimal
\textit{non-zero} distance to other handles to exclude those
connected handles. It is clear that the transformation at the
position of a handle $H_i$ is commonly determined by all handles
in $\mathcal{H}$ although the influence of far away handles is
trivial. On the other aspect, the smoothness of deformation is
improved to $C^{\infty}$. Cages can be formed by linking the
segment handles into closed loops. For example when editing the
portrait shown in Fig.\ref{figPortrait}, the cage located at the
boundary help resize the image. Moreover, the cage at the left eye
fully controls the shape inside it and therefore preserves the
salient feature.

%%%%%%%%%%%%%%%%%%%%%%%%%%%%%%%%%%%%%%%%%%%%%%%%%%%%%%%%%%%%%%%%%%%%%%%%%%
%%%%%%%%%%%%%%%%%%%%%%%%%%%%%%%%%%%%%%%%%%%%%%%%%%%%%%%%%%%%%%%%%%%%%%%%%%
%\begin{figure*}
%\includegraphics[width=\linewidth]{figLeaningTower}
%\caption{An example of processing the photograph of leaning tower
%by the cages. It can be found that the image can be easily
%warped with the help of cages and segment handles provided in our
%framework.}\label{figLeaningTower}
%\end{figure*}
\begin{figure}\centering
\includegraphics[width=\linewidth]{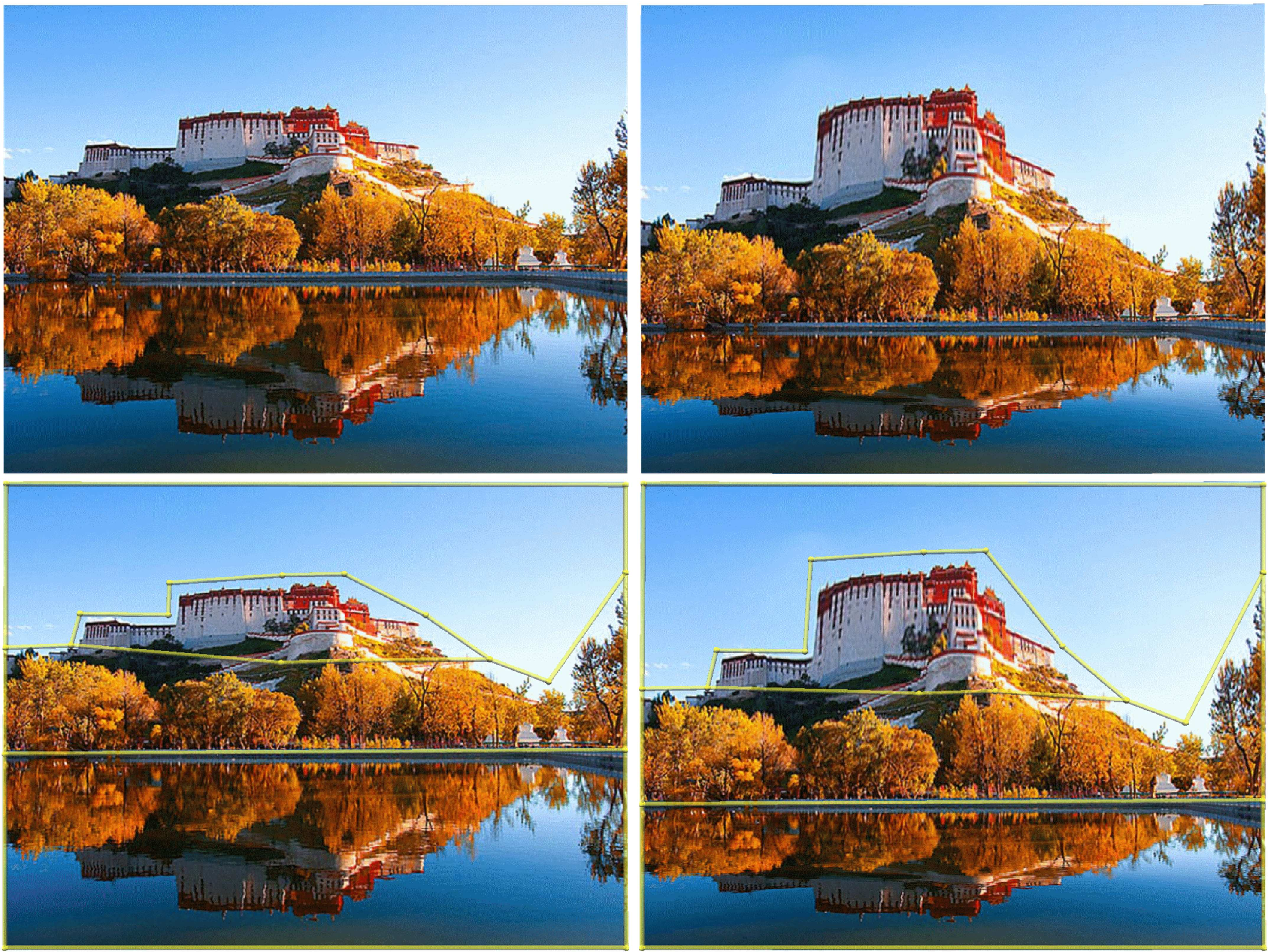}
\caption{An example of processing the photograph of Tibet palace
by segment handles. The image can be easily warped with the help
of segment handles provided in our framework.}\label{figPalace}
\end{figure}

\section{Results}\label{secResult}
Our meshfree weighting method provides a compact tool to assign
continuous weights for all points in the domain of deformation.
With the help of sophisticated techniques for assigning
transformations on the handles (e.g., the pseudo-edge method in
\cite{jacobson2011bounded} or the optimization method in
\cite{Jacobson2012AutoSkin}), a natural user interface for shape
deformation can be achieved.

%\begin{figure}[t]\centering
%\includegraphics[width=\linewidth]{figOctopus2D}
%\caption{An example to shown the simplicity of using point handles
%to drive the deformation on an Octopus. Note that, when only
%specifying translations at handles (upper-right), no rotation will
%be generated. This leads to a result with unreal distortion. The
%problem can be solved by using heuristic methods to add rotations
%on handles (the bottom row).}\label{figOctopus2D}
%\end{figure}
\begin{figure}\centering
\includegraphics[width=\linewidth]{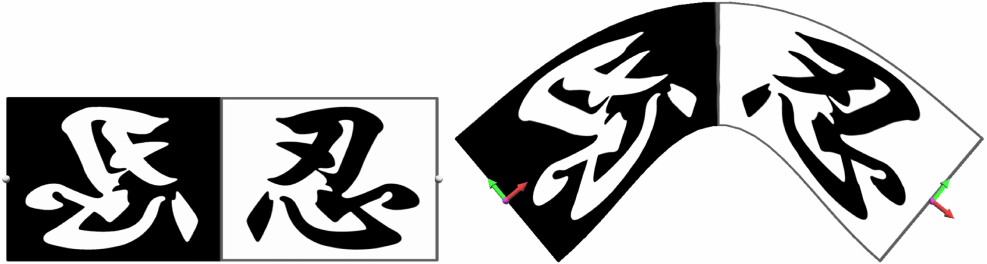}
\caption{An example of symmetric deformation: when applying
symmetric transformations on two symmetric handles to deform a
symmetric domain, our meshfree approach guarantees to obtain a
symmetric result.}\label{figChineseCharacter}
\end{figure}
We have tested this approach in a variety of examples by using
both the point and the segment handles. Figures \ref{fig:teaser},
\ref{figVirtualHandle} and \ref{figRabbit} have already
demonstrated the functionality of point handles. Especially, in
Fig.\ref{figVirtualHandle}, the scheme of virtual handles
insertion guarantees the interpolation at real handles. Figure
\ref{figRabbit} illustrates the effectiveness of our method in
determining transformations on virtual handles. The example of
using segment handles to deform a portrait has been shown in
Fig.\ref{figPortrait}. Another example is given in
Fig.\ref{figPalace} to warp the the shape of palace.
%The simplicity of using point handles to editing an octopus is illustrated in Fig.\ref{figOctopus2D}.
To obtain natural bending
results, we can add rotations on handles by heuristic methods
(e.g., the pseudo-edge \cite{jacobson2011bounded}). Another
example is to demonstrate the performance of our approach in a
symmetric deformation. When deforming a symmetric domain by adding
symmetric transformations on symmetric handles, it is expected to
get a symmetric result. This property is preserved by our
formulation (see Fig.\ref{figChineseCharacter}).

%\begin{figure}\centering
%\includegraphics[width=\linewidth]{figMonaLisa}
%\caption{A.}\label{figMonaLisa}
%\end{figure}

\begin{figure}[t]\centering
\includegraphics[width=\linewidth]{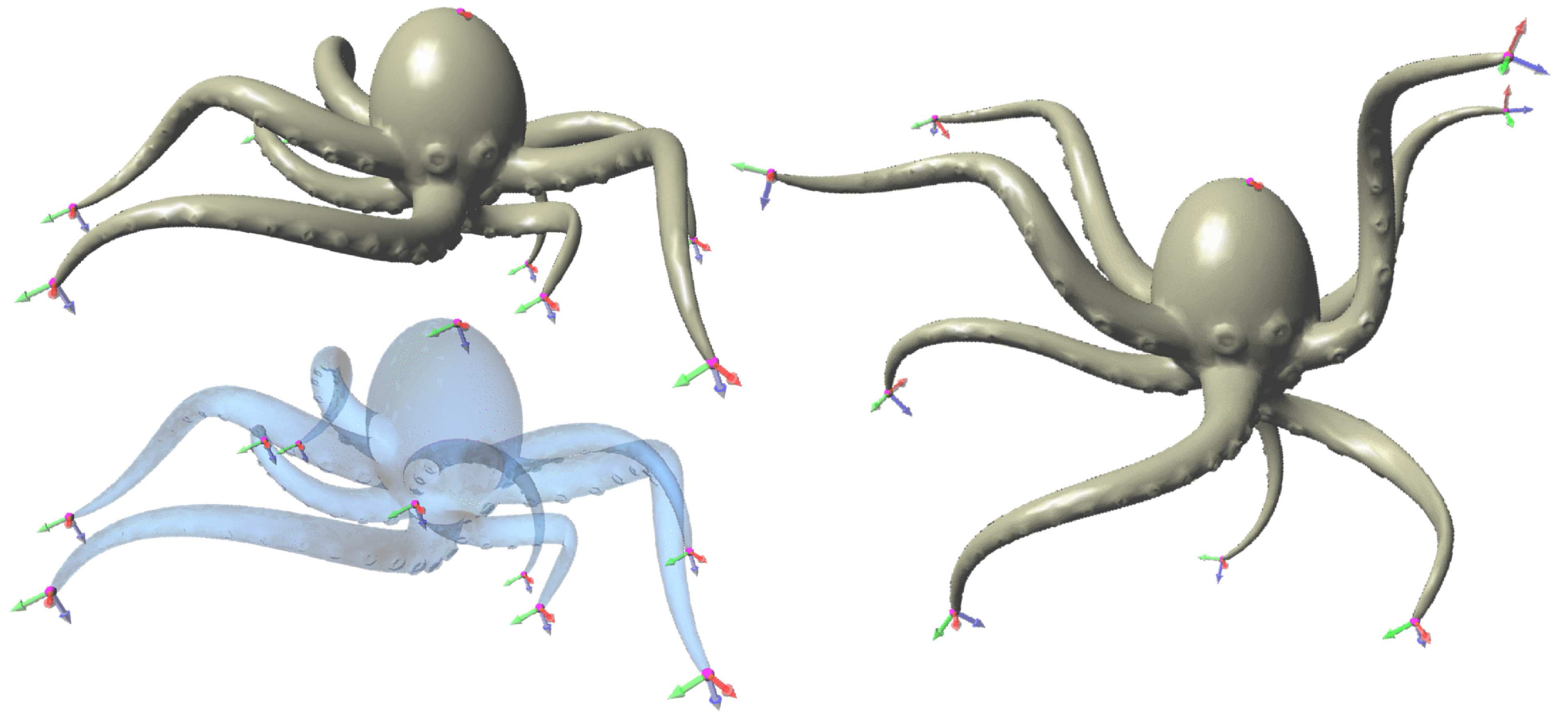}
\caption{The flexible 3D Octopus can be easily manipulated by
using the point handles.}\label{figOctopus3D}
\end{figure}

\begin{figure*}[!t]\centering
\includegraphics[width=\linewidth]{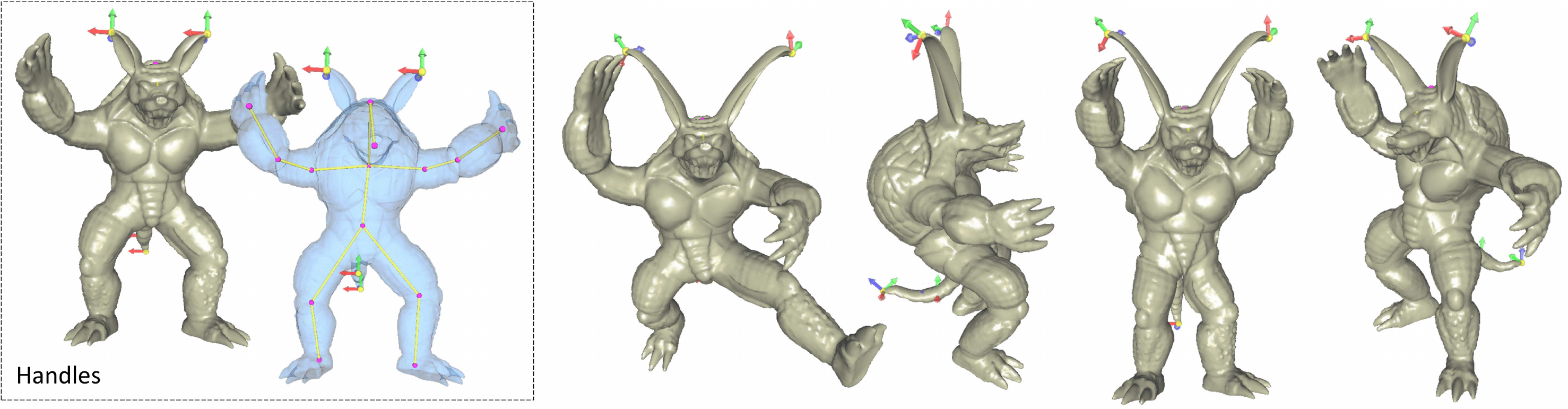}
\caption{An example of using the weights determined by our
approach in the animation of Armadillo driven by the point and the
segment handles.}\label{figArmadilloSkeletonDef}
\end{figure*}

\begin{figure*}[!t]\centering
\includegraphics[width=\linewidth]{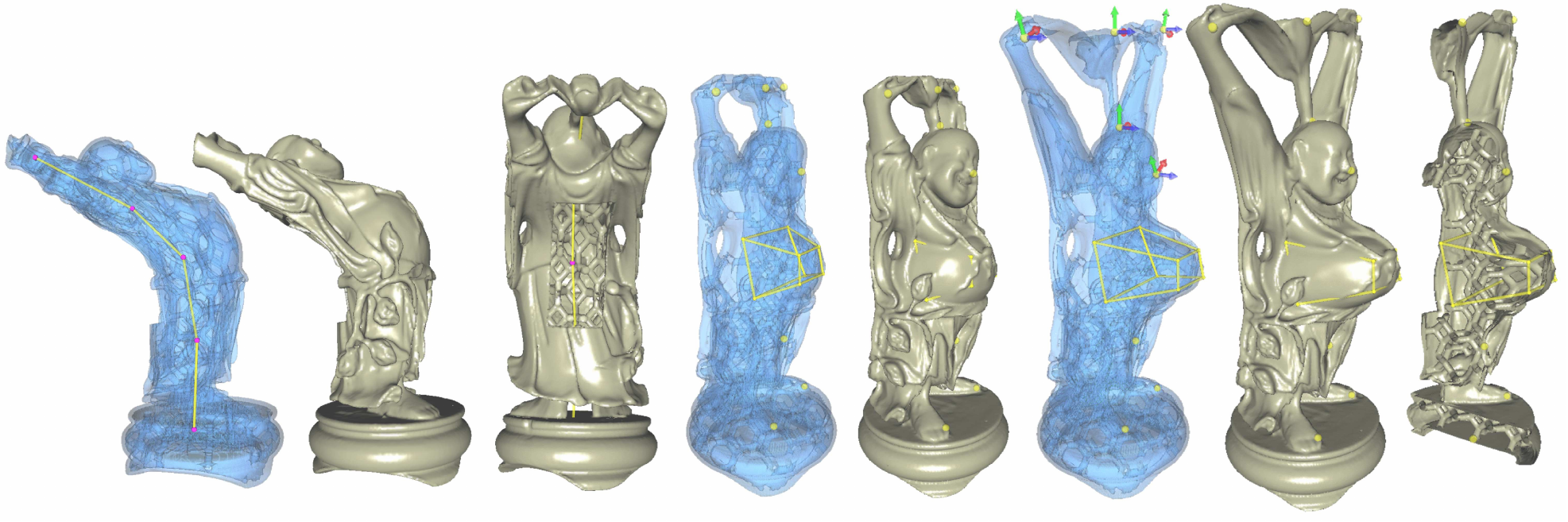}
\caption{An example of deforming a model with very complex
topology -- the Buddha model with interior truss structures, where
our meshfree approach can determine the weights for linear
blending effectively and efficiently.}\label{figBuddhaDef}
\end{figure*}

\begin{figure}[t]\centering
\includegraphics[width=\linewidth]{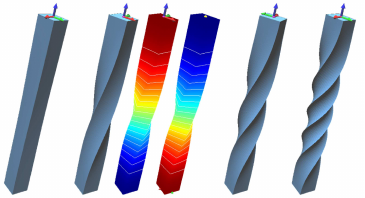}
\caption{An example of progressively twisting a bar with sharp
edges in different rotations: $\pi/4$, $\pi/2$ and $2 \pi$. The
color maps show the distribution of weights according to two
handles. The twists with large rotations are generated by
progressively applying small rotations -- e.g., $2^{\circ}$ per
update in our practice.}\label{figTwist}
\end{figure}

We also apply this method to deform 3D models. In these examples,
the 3D models are represented by polygonal mesh surfaces. The
weights computed by our approach are used in a linear blending way
to determine the new positions of vertices. Note that, the space
enclosed by a mesh surface need to be sampled into points with the
help of voxelization technique (e.g., \cite{Schwarz2010Voxel}) in
order to evaluate the discrete intrinsic-distance in the domain to
be deformed. Point handles are used to manipulate the flexible
Octopus in Fig.\ref{figOctopus3D}, where the interface of
manipulation becomes user-friendly after employing the scheme of
pseudo-edges to determine the transformation of point handles.
Linear blending scheme is widely employed in the animation of
skeletal models (e.g.,
\cite{Jacobson2011SA,Magnenat-Thalmann1989}). The example shown in
Fig.\ref{figArmadilloSkeletonDef} gives the performance of our
approach in this scenario. 3D models with very complex topology
(e.g., the Buddha model with internal truss structrues in
Fig.\ref{figBuddhaDef}) that are hard to be meshed can be easily
handled in our approach. When deformations with large rotation are
applied (e.g., in Fig.\ref{figTwist}), a progressive deformation
strategy can help generate satisfactory results.

\begin{table}\centering
\begin{tabular}{|l|c|c|c|c|}
  \hline
   & $|\mathcal{H}|$ & $|\mathcal{S}|$ & $t_{Vol}$ (sec.) & $t_{w}$ (sec.) \\
  \hline
  \hline
  Gingerman & 2 (8) & 155,457 & 0.584 & 0.054 \\
  Alligator & 3 (3) & 53,225 & 0.128 & 0.015 \\
  Rabbit & 4 (15) & 22,972 & 0.128 & 0.015 \\
  Portrait & 11 & 4,225 & 0.029 & 0.005 \\
  Palace & 22 & 4,225 & 0.041 & 0.003 \\
%  Octopus (2D) & - & - & - & - \\
  Chinese & 2 & 4,076 & 0.008 & 0.001 \\
  \hline
  \hline
  Octopus & 10 & 7,485 & 0.039 & 0.002 \\
  Armadillo & 17 & 26,002 & 0.100 & 0.014 \\
  Buddha & 4 & 236,661 & 0.302 & 0.051 \\
         & 20 & 236,661 & 0.834 & 0.160 \\
  Bar & 2 & 4,765 & 0.009 & 0.002 \\
  \hline
\end{tabular}
\caption{Computational Statistics for the examples shown in the
paper. $|\mathcal{H}|$ denotes the number of handles (the number
of virtual handles is shown in the bracket)
 and $|\mathcal{S}|$ represents the number of sample points used in the computation.
 The columns under $t_{Vol}$ and $t_{w}$ state the time  used
 in the computation of the voronoi diagram and the weights respectively.}\label{tabStatistics}
\end{table}

For prior mesh-based approaches, the numerical system must be
solved once more when new handles are inserted. In our meshfree
weighting formulation, the time cost of adding new handles is very
trivial as the weights are determined in a closed-form. Table
\ref{tabStatistics} lists the statistics of our approach on
different examples. All the tests are conducted on a computer with
Intel Core i7-3740QM CPU at 2.70GHz with 8GB memory, where our
current implementation only uses a single-core. All results of
deformation can be obtained at an interactive speed.

\textbf{Discussion.} When using the meshfree formulation presented
in the paper to deform real 2D/3D objects, sample points are
adopted as the medium for realizing the computation. The
error-bound of computation on this discrete representation is
guaranteed by the density of samples. However, during the process
of a sequence of deformations, the density of points could be
changed dramatically. In this sense, a dynamic up-sampling step
should be integrated in the framework to preserve the error-bound
of intrinsic-distance computation. The image editing applications
can be implemented by using either the super-sampling technique or
the texture mapping on a mesh. In our framework, the cost of
weight evaluation is trivial after resampling. The bottleneck is
the computation of intrinsic-distances on the sample points. Our
current implementation is based on the Dijkstra's algorithm.
However, this shortest path problem with multiple sources can be
computed in parallel on the system with many-cores
\cite{Rong2011CVT}, which can result in a significant speedup and
will be implemented in our future work.

\begin{figure}[t]\centering
\includegraphics[width=.9\linewidth]{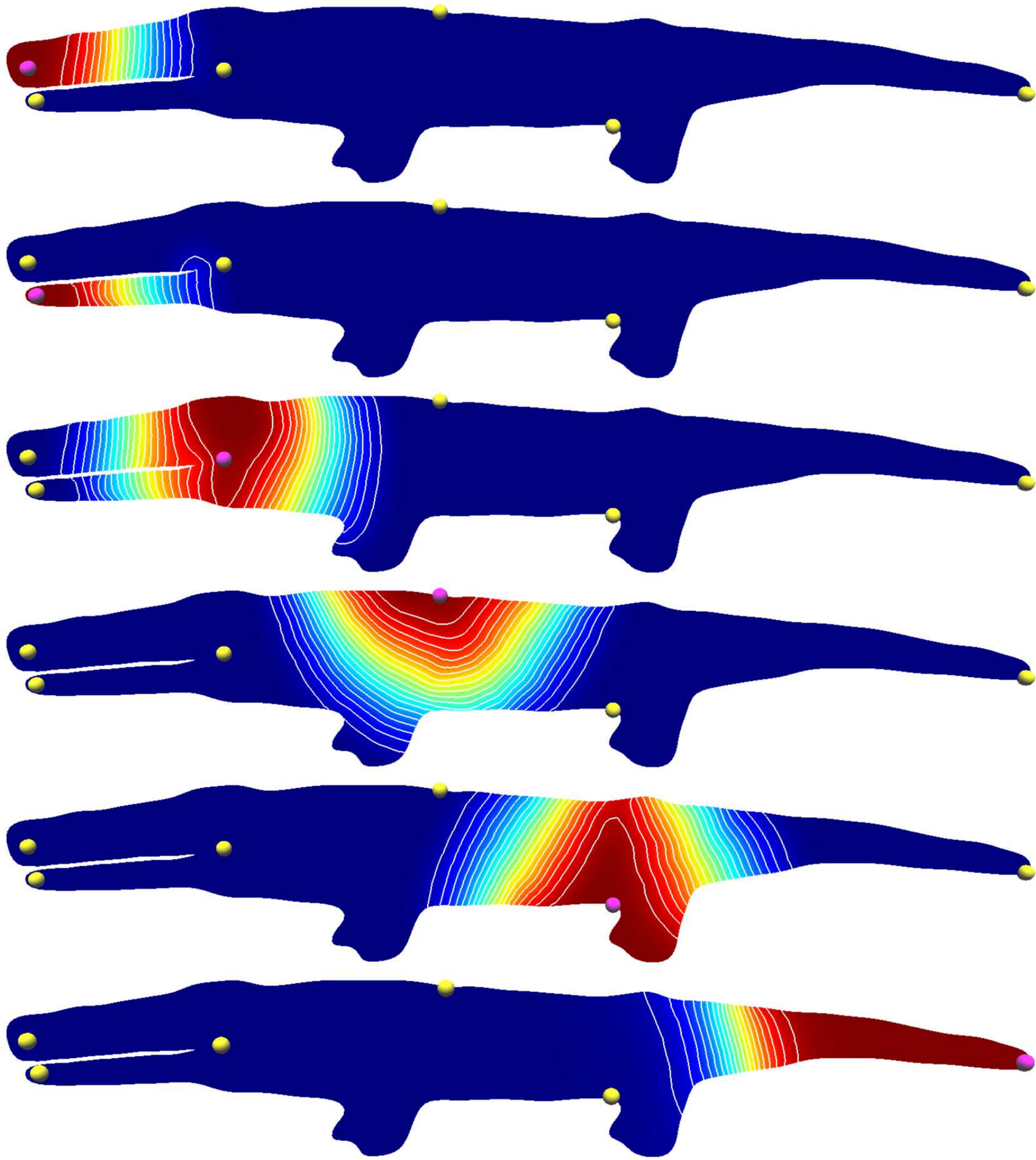}
\caption{The verification of no-local-maximum is taken by
analyzing the topology of isocurves on the weights' scaler-fields.
The handles (real and virtual) in this example are the ones shown
in Fig.\ref{figVirtualHandle}.}\label{figIsocurves}
\end{figure}

Our formulation gives global maximum at the positions of handles,
which is very important to avoid the unintuitive behavior of
deformations. For a shape-aware deformation, it is also demanded
having no-local-maximum. This has been verified in our
experimental tests. We check the topology of isocurves on the
fields of weights (see Fig.\ref{figIsocurves} for an example). If
there is a closed loop formed by isocurves of $w_i(\cdot)$ at one
place except the center of the handle $\mathbf{h}_i(\cdot)$, a
local maximum is generated there. However, no such case is found
in all our examples.

%%%%%%%%%%%%%%%%%%%%%%%%%%%%%%%%%%%%%%%%%%%%%%%%%%%%%%%%%%%%%%%%%%%%%%%%%%
%%%%%%%%%%%%%%%%%%%%%%%%%%%%%%%%%%%%%%%%%%%%%%%%%%%%%%%%%%%%%%%%%%%%%%%%%%
\section{Conclusion}\label{secConclusion}
We present a method to determine weights of blending for shape
deformation. Our formulation is meshfree and in a closed-form,
which can be easily used in a variety of applications in 2D/3D
deformations. Equipped with a virtual handle insertion algorithm,
good properties of weights generated by prior mesh-based methods
can all be preserved in this approach. A variety of examples have
been shown to demonstrate the function of our approach.

Only linear blending deformations are tested in the paper. We plan
to further extend the application of weights generated in this
approach to more advanced skinning methods, such as dual
quaternion \cite{Kavan2008DualQuaternion}, with which the blending
of two rigid motions will result in a rigid motion. This is a very
important property when the deformation of articulated characters
is computed by the skinning methods. The deformations driven by
linear blending are not always injective and therefore can
generate the results with foldovers and self-intersection.
Recently, some researches have been conducted in this direction to
produce injective mappings (e.g.,
\cite{Aigerman2013Injective,Schueller2013LIM}), which are mainly
mesh-based. In a function based formulation, the injectivity of a
mapping can be checked by the sign of Jacobian. However, it is
still not clear about how to resolve the problem when
self-intersection is detected. This will be one of our future
work.

%\begin{figure}[!ht]
%\centering
%  \includegraphics[width = 0.3 \columnwidth]{bar}\\
%  \caption{Bar deformation}\label{fig:bar}
%\end{figure}

%\section*{Acknowledgements}
%To Robert, for all the bagels.

\bibliographystyle{acmsiggraph}
\bibliography{SIGAsia14Deformation}

%%%%%%%%%%%%%%%%%%%%%%%%%%%%%%%%%%%%%%%%%%%%%%%%%%%%%%%%%%%%%%%%%%%%%%%%%%
%%%%%%%%%%%%%%%%%%%%%%%%%%%%%%%%%%%%%%%%%%%%%%%%%%%%%%%%%%%%%%%%%%%%%%%%%%
\section*{Appendix A: Intrinsic-Distance}\label{secSpaceDist}
For a 2-manifold shape in 2D/3D Euclidean space, all points on the
shape form a bounded domain $\Omega$. For any two points
$\{\mathbf{p}_s, \mathbf{p}_e\} \in \Omega$, if there exists a
curve line $\mathcal{C} \subset \Omega$ connecting $\mathbf{p}_s$
and $\mathbf{p}_e$, we define the intrinsic-distance of
$\{\mathbf{p}_s, \mathbf{p}_e\}$ along the curve $\mathcal{C}$ as
\begin{displaymath}
d(\mathbf{p}_s, \mathbf{p}_e; \mathcal{C})= length (\mathcal{C} )
\end{displaymath}
Then the intrinsic-distance of $\{\mathbf{p}_s, \mathbf{p}_e \}$
in $\Omega$ is defined as
\begin{displaymath}
d(\mathbf{p}_s, \mathbf{p}_e) = \min_{\mathcal{C}} \; length
(\mathcal{C} ).
\end{displaymath}
If there is no curve connecting $\mathbf{p}_s$ and $\mathbf{p}_e$,
that is the case they are not located in a connected region of
$\Omega$. The intrinsic-distance is then defined as
$d(\mathbf{p}_s, \mathbf{p}_e) = \infty$.

\textit{Sampling based intrinsic-distance.} \hspace{5pt} For a set
of sampling points $\mathcal{S} \in \Omega$ of $\Omega$, we can
build a graph $\mathcal{G}$ by using the sample points as nodes.
We represent the shortest distance between $\mathbf{p}_s$ and
$\mathbf{p}_e$ on $\mathcal{G}$ as $d_{\mathcal{G}}(\mathbf{p}_s,
\mathbf{p}_e; \mathcal{S})$. If for any two points
$\{\mathbf{p}_s, \mathbf{p}_e\} \in \Omega$, we always have
\begin{displaymath}
| d_{\mathcal{G}}(\mathbf{p}_s, \mathbf{p}_e; \mathcal{S}) -
d(\mathbf{p}_s, \mathbf{p}_e) | \leq \varepsilon,
\end{displaymath}
the sampling $\mathcal{S}$ is a \textit{distance-bounded sampling}
of $\Omega$.

The intrinsic-distance defined in this way has the following
properties:
\begin{itemize}
\item \textbf{Existence:} $d(\mathbf{p}_s, \mathbf{p}_e;
\mathcal{C})$ is always calculable once $\mathcal{C}$ is
determined, which is a curve segment in $\Omega$. Therefore,
$d(\mathbf{p}_s, \mathbf{p}_e)$ always exists for $\Omega$ when
$\mathbf{p}_s$ and $\mathbf{p}_e$ are located in the same
connected region.

\item \textbf{Uniqueness:} $d(\mathbf{p}_s, \mathbf{p}_e)$ is
uniquely determined while the corresponding curves may be
multiple.

\item \textbf{Convergency:} For any $\varepsilon > 0$, there
always exists an infinite sampling of $\Omega$ -- that is the
sampling density $D(\mathcal{S}) \rightarrow \infty$. Since
$\lim_{D(\mathcal{S}) \rightarrow \infty} \varepsilon = 0$, we
have
\begin{displaymath}
\lim_{D(\mathcal{S}) \rightarrow \infty}|
d_{\mathcal{G}}(\mathbf{p}_s, \mathbf{p}_e; \mathcal{S}) -
d(\mathbf{p}_s, \mathbf{p}_e) | = 0.
\end{displaymath}
\end{itemize}

%%%%%%%%%%%%%%%%%%%%%%%%%%%%%%%%%%%%%%%%%%%%%%%%%%%%%%%%%%%%%%%%%%%%%%%%%%
%%%%%%%%%%%%%%%%%%%%%%%%%%%%%%%%%%%%%%%%%%%%%%%%%%%%%%%%%%%%%%%%%%%%%%%%%%
\section*{Appendix B: Endpoint Constraints}\label{secEndPntConstraints}
From the analysis in \cite{Farin2001CAGD}, we know that
\begin{displaymath}
\mathbf{b}'(0)= n(\mathbf{b}_1 - \mathbf{b}_0), \quad
\mathbf{b}'(1)=n(\mathbf{b}_n - \mathbf{b}_{n-1})
\end{displaymath}
for a B\'{e}zier curve in $n$-th order. And also
\begin{displaymath}
  \mathbf{b}''(0) = n(n-1)(\mathbf{b}_2 - 2 \mathbf{b}_1 +
\mathbf{b}_0)
\end{displaymath}
\begin{displaymath}
  \mathbf{b}''(1) = n(n-1)(\mathbf{b}_n -
2\mathbf{b}_{n-1} + \mathbf{b}_{n-2})
\end{displaymath}
Incorporating the constraints in Eq.(\ref{eqConstraints}), we have
\begin{displaymath}
\mathbf{b}_1 = \mathbf{b}_0, \mathbf{b}_n = \mathbf{b}_{n-1},
\mathbf{b}_1 = \frac{\mathbf{b}_0+\mathbf{b}_2}{2},
\mathbf{b}_{n-1}=\frac{\mathbf{b}_n+\mathbf{b}_{n-2}}{2}.
\end{displaymath}
As we already need $\mathbf{b}_i^x=i/n$ to let $x=t$, it is not
difficult to find that
$\mathbf{b}_0^y=\mathbf{b}_1^y=\mathbf{b}_2^y=1$ and
$\mathbf{b}_n^y=\mathbf{b}_{n-1}^y=\mathbf{b}_{n-2}^y=0$ satisfy
all these constraints.

\end{document}